\documentclass[aip,reprint]{revtex4-1}

\usepackage[dvips]{graphicx}

\begin{document}


\title{Comment on ``Steady-state fluctuations of a genetic feedback
  loop: an exact solution'' [J. Chem. Phys. {\bf 137}, 035104 (2012).]}


\author{Guilherme C.P. Inoccentini}
\affiliation{Instituto de Matem\'atica e Estat\'{\i}stica,
  Universidade de S\~ao Paulo, Caixa Postal 66281, BR-05315-970 S\~ao
  Paulo, S.P., Brazil}
\author{Alexandre F. Ramos}
\affiliation{Escola de Artes, Ci\^encias e Humanidades, 
Universidade de S\~ao Paulo, 
Av. Arlindo B\'ettio, 1000 
CEP 03828-000, S\~ao Paulo, SP, Brazil}
\author{Jos{\'e} Eduardo M. Hornos}
\email{hornos@ifsc.usp.br}
\affiliation{Instituto de F\'{\i}sica de S\~ao Carlos,
 Universidade de S\~ao Paulo,
 Caixa Postal 369, BR-13560-970 S\~ao Carlos, S.P., Brazil}




\maketitle 


In the commented article modifications of the Spin-Boson Model
\cite{sasai03}, for a binary self-regulating gene, have been
proposed. The new master equations allow two different decay rates for
free and bound proteins. It also presents a vigorous critique of the
article 'Self-Regulated gene: an exact solution' \cite{hornos05} and
'subsequent publications'\cite{ramos07,schultz07,ramos10,ramos11}
reducing the finding of exact solutions for the spin model to ``just a
claim'' of the authors and not as a rigorous result with the strength
of a theorem.

Despite the main criticism on the mentioned solutions they are really
dueling with corrections to the master equations for stochastic gene
expression viewed as a many-body problem. In the section dedicated to
our article the exactness of our solution is questioned again and even
a demon, perhaps ironically, is used to explain one of the
approximations made in the formulation of the Spin Boson Model. The
task of the demon is the instantaneous replacement of the decaying
bound protein to the cytoplasm and the inexactness of the solution is
a heritage of the inexactness of the master equations.  The confusion
between solutions and equations goes until the end of the paper when
they state that 'our' equations (meaning the Spin Boson Formalism) and
theirs are dedicated to the same problem so one of them must be wrong!
Either demons or exact equations do not belong to science but the
concept of an exact solution for a differential equation is well
defined and exhaustively studied in the field of symmetries and
integrable systems. Indeed we have been able to show that solubility
of the Spin Boson models has underlying symmetries
\cite{ramos07,ramos11}.

Both equations contains several approximations and are very
far from biological details even if we considered a simpler induced
transcription in E.Coli. The most important limitation of this class of models  is the bypassing
of the whole transcription process and even a demon is not able to
take the information from the DNA, to build the mRNA and translate it
in the ribosomes instantaneously, putting back the protein to control
the promoter site. Using the terminology introduced  by the authors, science has a
plenty of demons and the crucial question is the selection of proper approximations needed
to predict and to explain the experiments successfully.

 In section V the authors presents a section named: Numerical Validation of The Exact Solution
which completes the confusion. Exact solutions, by definition, don't need a numerical check and are verified analytically. A solution can not be validated by a numerical simulation or even by an experiment: the occurrence of an agreement only says that the equations are not wrong. In contrary, the exact analytical solutions are frequently used to validate numerical simulations, stability of algorithms and so on.

The authors report the finding of an exact solution for the
steady-state genetic feedback loop. Examining the paper we could not
find a complete exact solution for the model as announced in the title, but just half of
the work has been carried out.  In fact, they present an exact general solution
for  one  probability distribution named $G_1(z)$ but they fail to get the other one  $G_0(z)$ in closed form as they recognize explicitly.
Mentioning that the equation obtained for this component is not of
Riemann type they only write a first order differential equation for
it, complaining that ``It is difficult to extract an explicit solution
for $G_0(z)$ by integrating this equation''. If the authors have read Refs. \cite{hornos05,ramos11} carefully, would learn that these equations are not Riemann type but from another family, the Heun equations, and therefore, still integrable.

Consequently the
normalization constant is not calculated obstructing the analyticity of
the whole model.  The underlying reason is that ``apparently unknown''
integrals are needed to obtain the exact solution. They also recognize
that they cannot calculate the fluctuations as a function of the
parameters of the model, without the numerical computation of the
normalization, but only the fluctuation divided by the mean value, also
known as the Fano factor. 

The use the generating function technique replaces the traditional recursive form of the master equations by partial differential equations with the introduction of the complex functions  $G_0(z) =
\sum_{n=0}^\infty P_0(n)z^n$ and $G_1(z) = \sum_{n=0}^\infty
P_1(n)z^n,$ where $n$ is the stochastic variable, the number of protein molecules,
and  $ P_0(n)$ and $P_1(n)$ are the usual probabilities. At steady state limit the problem reduces to the solution of the system
\begin{equation}\label{edo1}
(z-1)(\rho_{u}G_{0}-G_0') +
(\theta+\sigma_{u}z)G_{1}-\sigma_{b}zG_0' = 0,
\end{equation}
\begin{equation}\label{edo2}
(z-1)(\rho_{b} G_{1}-G_1') - 
(\theta+\sigma_{u}\ \ )G_{1}+\sigma_{b}\,G_0'=0.
\end{equation}
$G_1(z)$ may be written as a product between $A {\rm
  e}^{\rho_b(z-1)}$ and the KummerM function ${\rm M} (\alpha, \beta, w),$ where
\begin{eqnarray}
\alpha &=& \theta + \sigma_u \frac{\rho_u-\rho_b}{\rho_u - \rho_b - \rho_b \sigma_b}, \\ 
\beta &=& 1 + \theta +
\frac{\sigma_u + \sigma_b (\sigma_u + \rho_u) }{(1+\sigma_b)^3}, \\ 
w &=& (\rho_u - \rho_b - \rho_b \sigma_b) \frac{ (1+\sigma_b) z -1 } {(1+\sigma_b)^2}.
\end{eqnarray}

We obtain $G_0(z)$ taking linear combinations of the
Eqs. (\ref{edo1}) and (\ref{edo2}):
\begin{eqnarray}
G_0(z) = A {\rm e}^{\rho_b(z-1)} \times \left[
  \frac{1+\sigma_b}{\sigma_b}\,\frac{\alpha}{\rho_u}\, {\rm M}(\alpha+1,
  \beta, w) \right.  \nonumber \\ \left. 
   +\, \frac{\theta -
    \alpha}{\rho_u-\rho_b} \ \ {\rm M}(\alpha, \beta, w) \ \ \right],
\end{eqnarray}
as we can verify by direct substitution. Normalization constant $A$ follows from probability conservation as usual
and gives to this model same status of that presented by Hornos {\em
  et.al}.

Throughout their article the authors explain why our solutions are not
exact and theirs are. The reason is that the equations for the spin
boson model of Ref. \cite{hornos05} are not exact, some terms are
missing and others are ``non-physical''. Consequently our solution is
not exact because our equations are not exact. In contra-position
their stochastic equations are ``exact'' and therefore the partial
solution they presented are exact.

Even though we are not claiming that our solutions have the same
status, let us consider the exact solution for the Schr\"odinger
equation for hydrogen and numerical solutions for the Helium atom. Of
course the Schr\"odinger equation for an electron in the Coulomb field
is not exact and unphysical if one want to use those terms. The
Coulomb potential depends only on the radius, relativistic terms are
missing, quantized electromagnetic field is absent. The use of the
space dependent potential violates special relativity and is
incompatible with Maxwell equations that are Lorentz invariant. An
static potential between to charges is only possible if we assume the
existence of a "daemon" which instantaneously tells one charge that
the other did a small movement. Following their reasoning the
hydrogenic solution in terms of spherical harmonics and Laguerre
functions are not exact!  Of course this is not the case for two
reasons:({\em i}) there are no exact equations in Science and ({\em
  ii}) we consider an exact solution for a given equation if we can
solve it analytically in a closed form. Even if one relax the
definition of exact solution in closed form to numerical evaluation of
exact solution, {\em i.e.} numerical solutions to a target equation,
without approximations, calculated with desired precision, one cannot
call the exact solutions presented at Refs. \cite{hornos05,ramos11} as
just a claim or use quotes to diminish the value of the
calculation. In practice the exact solutions are powerful because they
allow any calculation and the study of the parametric equation of one
model. Furthermore, approximations and physical intuition are also
easier if we have exact solutions.

The stochastic equations proposed in their article are different from
the spin boson ones. Indeed there is no parameter that can be used to
obtain rigorously one set of equations from the other. Both are
unphysical and approximated. The transcription process, the ribosomes,
tRNA's and the amplification of the protein number by repeated
translation of one mRNA have been eliminated in both models which is a
radical oversimplification of any protein synthesis process. Even in
the transcription of the {\em lac} operon of E.Coli controlled by the
{\em Lac Repressor}, the comparison with experiments requires two
stochastic processes \cite{Thattai01}. In the compared models another
daemon is required, it carries one mRNA on a hand and the ribosomes on
the other, translating instantaneously and repeatedly the anticodon
information.

The proposed model is welcome and has some
interesting modifications as claimed by the authors. The
instantaneous reposition of protein to the binding mode is avoided
and the protein degradation when the protein is bound is taken into
account, even though a new parameter must be included. However, the strict
interpretation of both models, and, consequently, of the parameters,
is not promising.  The transcription and translation processes
considered here have several time scales. For example, the free
proteins degradation, the bacterial division time, the mRNA
degradation time that usually is much smaller than the protein
degradation, the discrepancy between mRNA and protein number of
molecules by more than an order of magnitude in E.Coli
\cite{cai06}. This means that we must have an extra guiding principle
to perform a physical analysis and the best one are the available experiments. The critique present here do not
diminishes the importance of their proposal which can be
considered an alternative for negative self-regulation.



%

\end{document}